\documentclass[11pt]{article}
  \usepackage{amssymb}

     \renewcommand{\baselinestretch}{1.3}
  \renewcommand{\arraystretch}{1.2}
\begin{document}

 \title{Ordered Probability Mass Function}

 \author{Zhengjun Cao$^1$, \qquad Lihua Liu$^2$\\
 {\small $^1$Department of Mathematics, Shanghai University,
  China. \textsf{caozhj\,@\,shu.edu.cn}}\\
 {\small $^2$Department of Mathematics, Shanghai Maritime University, Shanghai,
  China.}  }

 \date{}\maketitle

\begin{quote}
\textbf{Abstract} Suppose that in the four  tests  Alice's  scores  are 90, 95, 85, 90, and  Bob's scores are 85, 95, 90, 90.
How to evaluate their scores? In this paper, we introduce the concept of ordered probability mass function which can be used to find a probability mass function with smaller variance. More interestingly, we can use it to distinguish  sequences of positive numbers statistically.

\end{quote}

 \section{Introduction}
 A random variable that can take on at most a countable number of possible values is said to be discrete.
Suppose that $X$ is a discrete random variable whose possible values are $x_1, x_2, \cdots, x_n$, $x_i\geq 0, i=1, \cdots, n$. We call the probability mass function   $$ p(x_i)=P(X=x_i)=1/n, \ i=1, \cdots, n $$ as average probability mass  function (APM for short),
 the probability mass function  $$ p(x_i)=P(X=x_i)=\frac{x_i}{\sum_{i=1}^n x_i}, \ i=1, \cdots, n  $$ as general probability mass  function (GPM).  In a typical statistical problem, we have a random variable $X$ of interest, but its probability mass function is not known.
 The above two probability mass functions are widely used in daily life.

 The quantity $E[X]$ defined by $\sum_{x: p(x)>0} x p(x) $ is called the expected value of $X$, which is also called the \emph{expectation} of $X$.  The variance \cite{HMC12,R10}, which is equal to the expected square of the difference between $X$ and its expected value, is a measure of the spread of the possible values of $X$. Practically speaking, a probability mass function with all positive probabilities and smaller variance is more appreciated.

    How to find a  probability mass function with  smaller variance? We have observed that there were no methods which can be used to find it.
   We have also observed that there were no methods which can distinguish  sequences of positive numbers statistically, because
   the common sample mean and sample variance do not depend on the order relation of elements in a sample space.
   In this paper, we introduce the concept of ordered probability mass function which can be used to find a probability mass function with smaller variance.
    More interestingly, it can be used to distinguish  sequences of positive numbers statistically.
     To the best of our knowledge, it is the first time to introduce a method to cope with this problem.

 \section{Ordered probability mass function}

In order to introduce the new concept of ordered probability mass function, we prove the following lemma.

 \textbf{Lemma 1}. \emph{If $abc=1$, then
 $$\frac{a}{1+a+ab}+\frac{b}{1+b+bc} +\frac{c}{1+c+ca}=1 \eqno(1) $$}

 \emph{Proof}. Since $abc=1$, we have
  $$\frac{b}{1+b+bc}=\frac{ab}{a+ab+abc}=\frac{ab}{1+a+ab},$$
 $$\frac{c}{1+c+ca}=\frac{abc}{ab+abc+abca}=\frac{1}{1+a+ab}.$$
  Thus Eq.(1) holds.  \hfill $\Box$

  Likewise, if $abcd=1$, then
  $$\frac{a}{1+a+ab+abc}+\frac{b}{1+b+bc+bcd} +\frac{c}{1+c+cd+cda}+\frac{d}{1+d+da+dab}=1 \eqno(2) $$

More generally, if $x_1x_2\cdots x_n=1$, then
 $$\sum_{i=1}^{n}\frac{x_i}{1+x_i+x_ix_{i+1}+\cdots+ x_i\cdots x_n+ x_i\cdots x_nx_1+\cdots+ x_i\cdots x_nx_1\cdots x_{i-2}}=1 \eqno(3)$$

Suppose that $\frac{x_i}{1+x_i+x_ix_{i+1}+\cdots+ x_i\cdots x_n+ x_i\cdots x_nx_1+\cdots+ x_i\cdots x_nx_1\cdots x_{i-2}}> 0, i=1, \cdots, n, $ then  Eq.(3) provides us a new probability mass function.

\textbf{Definition 1.}
Suppose that  $X$ is a discrete random variable whose possible values are $x_1, x_2, \cdots, x_n$, $x_i> 0, i=1, \cdots, n$.   We call
  the probability mass function  \begin{eqnarray*} & & p(x_i)=P(X=x_i)\\
&=& \frac{x_i}{1+x_i+x_ix_{i+1}+\cdots+ x_i\cdots x_n+ x_i\cdots x_nx_1+\cdots+ x_i\cdots x_nx_1\cdots x_{i-2}} , \ i=1, \cdots, n  \quad (4)
\end{eqnarray*}  as ordered probability mass function (OPM) provided that $x_1x_2\cdots x_n=1$.

By the  definition, the OPM for the sequence $ \frac 5 6, 1, 1, \frac 6 5$ is computed as follows:
\begin{eqnarray*}
  p(\frac 5 6)& =& \frac{\frac 5 6}{1+\frac 5 6+\frac 5 6\times 1+ \frac 5 6\times 1\times 1}= \frac 5 {21},\\
 p(1)&=& \frac{1}{1+1+1\times 1+   1\times 1\times \frac 6 5}= \frac 5 {21}, \quad \mbox{(for the first element ``1" in the sequence)}\\
   p(1)&=& \frac{1}{1+1+1\times \frac 6 5+    1\times \frac 6 5\times \frac 5 6}= \frac 5 {21},\quad \mbox{(for the second element ``1" in the sequence)}\\
 p(\frac 6 5)&=&\frac{\frac 6 5}{1+\frac 6 5+ \frac 6 5 \times \frac 5 6+  \frac 6 5 \times \frac 5 6\times 1}= \frac 6 {21}.
 \end{eqnarray*}

It is easy to find that the  APM and GPM do not depend on the order relation of elements in a sample space, but the OPM depends on it. Here is an illustration example.

\textbf{Example 1}. Find the APM, GPM, OPM, expectations and variances of the following sequences.

\begin{center}
 \begin{tabular}{|l|l|c|c|}
   \hline
sequence &  probability mass &  Expectation & Variance \\ \hline
     $  \frac 6 5, 1, 1, \frac 5 6  $& APM: $\frac {1}{4}, \frac {1}{4}, \frac {1}{4}, \frac {1}{4}$  & $ 1.00833 $ & 0.0168750 \\ \hline
   $ \frac 6 5, 1, 1, \frac 5 6  $ & GPM: $\frac{36}{121},  \frac{30}{121}, \frac{30}{121}, \frac{25}{121} $ & $ 1.02507 $  & $0.0170116$\\
   \hline
   $ \frac 6 5,\frac 5 6, 1, 1$  & OPM: $ \frac{6}{21},\frac{5}{21}, \frac{5}{21}, \frac{5}{21}$ & $ 1.01746$ & $0.0177375$\\
   $ \frac 6 5, 1, \frac 5 6,  1$  & OPM: $  \frac{6}{22}, \frac{6}{22}, \frac{5}{22}, \frac{5}{22}$ & $ 1.01667$ & $0.0169444$\\
  $  \frac 6 5,  1, 1, \frac 5 6 $ & OPM: $  \frac{6}{23}, \frac{6}{23}, \frac{6}{23}, \frac{5}{23}$ & $1.01594$ & $\underline{0.0162193}$\\ \hline
 \end{tabular}\end{center}
 Note that the sequence $  \frac 6 5,  1, 1, \frac 5 6,$ with ordered probability mass $ \frac{6}{23}, \frac{6}{23}, \frac{6}{23}, \frac{5}{23}$, has the smallest variance.
From the practical point of view, the probability distribution  $$ \left\{\left(\frac 6 5; \frac{6}{23}\right), \left(1; \frac{6}{23}\right), \left(1; \frac{6}{23}\right), \left(\frac 5 6; \frac{5}{23}\right)\right\}$$ is more appreciated.

\textbf{Example 2}. Compute the APM, GPM, OPM, expectations and variances of the following sequences.
\begin{center}
 \begin{tabular}{|l|l|c|c|}
   \hline
sequence &  probability mass &  Expectation & Variance \\ \hline
     $  \frac 6 5, \frac 7 6, \frac 6 7, \frac 5 6  $& APM: $\frac {1}{4}, \frac {1}{4}, \frac {1}{4}, \frac {1}{4}$  & $ 1.01429 $ & 0.0287868 \\ \hline
   $ \frac 6 5, \frac 7 6, \frac 6 7, \frac 5 6   $ & GPM: $\frac{252}{852},  \frac{245}{852},  \frac{180}{852}, \frac{175}{852} $ & $ 1.04267 $  & $\underline{0.0280154}$\\
   \hline
  $ \frac 6 5, \frac 7 6, \frac 6 7, \frac 5 6   $ & OPM: $ \frac{6}{24},\frac{7}{24}, \frac{6}{24}, \frac{5}{24}$ & $ 1.02817$ & $0.0281971$\\
   $ \frac 6 5,  \frac 7 6, \frac 5 6,   \frac 6 7  $ & OPM: $  \frac{36}{143}, \frac{42}{143}, \frac{35}{143}, \frac{30}{143}$ & $1.02854$ & $0.0284942$\\
   $ \frac 6 5,  \frac 6 7, \frac 7 6, \frac 5 6   $ & OPM: $  \frac{42}{155}, \frac{36}{155}, \frac{42}{155}, \frac{35}{155}$ & $ 1.02854$ & $0.0285634$\\
  $ \frac 6 5,  \frac 6 7,  \frac 5 6, \frac 7 6   $ & OPM: $  \frac{42}{143}, \frac{36}{143}, \frac{30}{143}, \frac{35}{143}$ & $1.02860$ & $0.0286940$\\
    $ \frac 6 5,  \frac 5 6,  \frac 6 7, \frac 7 6   $ & OPM: $  \frac{42}{142}, \frac{35}{142}, \frac{30}{142}, \frac{35}{142}$ & $1.02897$ & $0.0289964$\\
     $ \frac 6 5,  \frac 5 6,  \frac 7 6, \frac 6 7   $ & OPM: $  \frac{36}{131}, \frac{30}{131}, \frac{35}{131}, \frac{30}{131}$ & $1.02861$ & $0.0286305$\\ \hline
 \end{tabular}\end{center}
Note that  $ \frac 6 5, \frac 7 6, \frac 6 7, \frac 5 6   $ with the general probability mass  $\frac{252}{852},  \frac{245}{852},  \frac{180}{852}, \frac{175}{852}, $ has
the smallest variance.

\section{Application}

As we mentioned before, a probability mass function with smaller variance is more appreciated in practice. Equipped with the tool of ordered probability function,
we now can handle  the following problem:  In the four  tests,  Alice's  scores  are 90, 95, 85, 90, and  Bob's scores are 85, 95, 90, 90.
How to evaluate their scores?

To meet the requirement that $x_1x_2\cdots x_n=1$ in the definition of OPM, we have first to standardize their scores.
 Set $\mathrm{v}=\sqrt[4]{90\times 95\times 85\times 90}$, $a_1=\frac{90}{\mathrm{v}}, a_2=\frac{95}{\mathrm{v}}, a_3=\frac{85}{\mathrm{v}}, a_4=\frac{90}{\mathrm{v}},$  $b_1=\frac{85}{\mathrm{v}}, b_2=\frac{95}{\mathrm{v}}, b_3=\frac{90}{\mathrm{v}}, b_4=\frac{90}{\mathrm{v}}.$ Hence, $a_1a_2a_3a_4=1,$ $b_1b_2b_3b_4=1.$ We then compute the ordered probability mass functions of sequences $a_1, a_2, a_3, a_4,$ and $b_1, b_2, b_3, b_4.$

\begin{center}
 \begin{tabular}{|l|l|c|c|c|}
   \hline
& sequence & OPM &  Expectation & Variance \\ \hline
 Alice &    $  a_1, a_2, a_3, a_4  $& 0.246667, 0.260572, 0.246286, 0.246476 & 1.00157 & 0.00156616 \\ \hline
Bob&    $ b_1, b_2, b_3, b_4 $ &  0.239718, 0.253231, 0.253427, 0.253623 & 1.00152 & 0.00152324\\
   \hline
   \end{tabular}\end{center}

   Since Bob's ordered probability mass function has smaller variance 0.00152324, we claim that Bob wins the four tests.

   \section{Conclusion}
   The order relation of elements in a sample space is  ignored
   because the common sample mean and variance do not depend on the relation. Thus the common statistical methods can not evaluate the sequence
   90, 95, 85, 90, and the sequence  85, 95, 90, 90.
   The proposed concept of ordered probability function can be used to distinguish sequences of positive numbers.
   We believe that the novel method developed in this paper will have more practical applications in the future.

\end{document}